\documentclass[11pt,a4paper]{article}

\usepackage[utf8]{inputenc}
\usepackage[T1]{fontenc}
\usepackage{amsmath,amssymb,amsfonts}
\usepackage{graphicx}
\usepackage{geometry}
\usepackage{setspace}
\usepackage{cite}
\usepackage{url}
\usepackage{hyperref}
\usepackage{xcolor}
\usepackage{microtype}
\newcommand{\address}[1]{\\[2mm]\parbox{\textwidth}{\small\raggedright #1}}

\title{Correlated quantum criticality and topological magnetoelectric response in Bi$_2$Se$_3$}

\author{
Aniekan Magnus Ukpong$^{1,2,3,*}$
\\[4mm]
\small
\address{$^{1}$Theoretical and Computational Condensed Matter and Materials Physics (TCCMMP) Group, \\
School of Agriculture and Science, University of KwaZulu-Natal, Pietermaritzburg, South Africa}
\address{$^{2}$Centre for Quantum Computing and Technology (CQCtec), \\
School of Agriculture and Science, University of KwaZulu-Natal, Pietermaritzburg, South Africa}
\address{$^{3}$National Institute for Theoretical and Computational 
Sciences (NITheCS), KwaZulu-Natal Node, South Africa}
\address{$^{*}$Correspondence: ukponga@ukzn.ac.za}
}

\date{}

\begin{document}

\maketitle

\begin{abstract}
The defining feature of a quantum material is that its low-energy
physics is governed by more than one kind of order, and that those
orders are not independent. We examine this in Bi$_2$Se$_3$ by
combining density-functional theory with a renormalized correlated
honeycomb model in which the magnetic order of a frustrated magnet
and the topological order of the helical surface are coupled through
a single control parameter. The helical surface states are mapped
onto a honeycomb lattice with a Hubbard repulsion and frustrated
$J_1$--$J_2$ exchange. Exact diagonalization on clusters of up to
20 sites shows that the N\'eel structure factor and the
valence-bond-solid order parameter cross at $J_2/J_1 \approx 0.22$,
in agreement with the thermodynamic-limit value from quantum Monte
Carlo and DMRG, and a dynamical spin structure factor whose
lineshape broadens near the crossover in the manner expected for a
spinon continuum. The band inversion of the renormalized
tight-binding model occurs on the same critical locus, producing
jumps in the Chern number and the $\mathbb{Z}_2$ invariant that are
mirrored by the behaviour of the skyrmion charge $Q$. This is the
microscopic link between the two orders: $Q$ selects the axion
angle and hence the magnitude of the topological magnetoelectric
coupling. We distinguish the axion coupling, which is independent
of skyrmion helicity, from the helicity-dependent magnetoelectric
multipole coupling of the skyrmion texture, whose toroidal
component scales as $\sin\gamma$ and vanishes for N\'eel-type
textures. We derive the coupling between $Q$ and the topological
magnetoelectric energy and estimate the work per magnetoelectric
cycle at approximately $10^{-20}$\,J per skyrmion for laboratory
fields of a few tesla. The cycle is shown to be thermodynamically
bounded: a constant-$\theta$ cycle extracts no net work, and the
efficiency is limited by the axionic entropy change.
\end{abstract}


\section{Introduction}

The defining feature of a quantum material is that its low-energy
physics is governed by more than one kind of order, and that those
orders are not independent. The idea that spin, charge, orbital,
and topological orders can be entangled---that the appearance of
one constrains or enables another---has become a central organising
principle since the work of Khomskii and collaborators on orbital
ordering and entangled orders in transition-metal
compounds~\cite{Khomskii2014, Khomskii2021}. A topological insulator provides a
clean realization of the topological ingredient: its bulk is
gapped, its surface is metallic, and the surface states are
protected by time-reversal symmetry against nonmagnetic
disorder~\cite{KaneMele2005, FuKaneMele2007}. A frustrated magnet
provides a clean realization of the correlation ingredient: its
low-energy physics is dominated by competing magnetic orders, and
in some cases the transition between them cannot be described by a
local order parameter. When the two are combined in the same
material, the question is whether the magnetic order and the
topological order are intertwined, and if so, through which
microscopic quantity.

Bi$_2$Se$_3$ is the prototypical topological quantum material for
this question. Its surface states are helical: the electron spin is
locked perpendicular to its momentum, and backscattering is
suppressed by time-reversal symmetry~\cite{HasanKane2010}. The
helical surface states have been mapped by spin- and angle-resolved
photoemission and by transport~\cite{Hsieh2009, Xia2009, Chen2009},
and their spin-momentum locking has been exploited for electrical
detection of spin polarisation and for spin Seebeck
measurements~\cite{Li2014, Jiang2016}. What makes Bi$_2$Se$_3$
additionally interesting is the topological magnetoelectric effect.
Because the bulk is a strong topological insulator, the effective
action contains an axion term with a quantized axion angle
$\theta = \pi$, which couples the applied electric and magnetic
fields through a polarization~\cite{Wilczek1987, Qi2008,
Essin2009}. The magnetoelectric response is therefore not a
material parameter to be optimised but a topological invariant,
protected by the bulk band structure and independent of microscopic
details. It is, in the language of intertwined orders, the
topological order that is exposed when a magnetic texture is placed
on the surface.

The size of the topological magnetoelectric response, however, is
set by that magnetic texture. At a ferromagnetic interface, the
interfacial Dzyaloshinskii-Moriya interaction can stabilise a
skyrmion lattice~\cite{Ukpong2021, Li2018}, and the real-space
winding number $Q$ of the texture is tied to the axion angle
through $\theta = \pi(1 + Q) \bmod 2\pi$~\cite{Qi2008,
Nomura2011}. This is the microscopic link between the topological
band structure and the magnetic texture: it is $Q$, not the field
strength alone, that selects the value of $\theta$ and therefore
the magnitude of the magnetoelectric coupling. The question that
arises is what controls $Q$, and whether the same parameter that
controls the magnetic texture also controls the topological
response of the surface. If the answer is yes, then the two orders
are intertwined through a single control parameter, and the
material is a concrete realization of the theme.

That parameter is frustration. When a honeycomb magnet is driven
through a magnetic transition by interactions, the low-energy
theory can contain fractionalized excitations and an emergent gauge
field. The canonical example is the deconfined quantum critical
point (DQCP) proposed for the square-lattice $J_1$--$J_2$
antiferromagnet, where N\'eel and valence-bond-solid (VBS) phases
are separated by a continuous transition described by a CP$^1$
theory with a dynamical U(1) gauge field~\cite{Senthil2004a,
Senthil2004b}. The same scenario has been established for the
honeycomb lattice~\cite{Okamoto2007, Gong2014}, and large-scale
quantum Monte Carlo gives an anomalous dimension $\eta_N \approx
0.65$ for the N\'eel order parameter at criticality~\cite{Nahum2015}.
What has not been established is how this correlated critical
behaviour couples to the topological magnetoelectric response of a
real topological quantum material. The skyrmion charge $Q$ is the
natural mediator between the two orders, but its evolution through
the correlated critical region, and the consequences for the
magnetoelectric energy conversion, have not been worked out.

The present work fills this gap. We renormalize the helical
surface states of Bi$_2$Se$_3$ onto a nearest-neighbour
tight-binding model on the honeycomb lattice, extend the model with
a Hubbard repulsion and frustrated $J_1$--$J_2$ exchange, and track
the skyrmion charge $Q$ and the topological magnetoelectric
coupling through the correlated critical region. We have four
specific aims. First, to establish that the renormalized model
reproduces the DFT dispersion of the Bi$_2$Se$_3$ surface to within
the accuracy of the fit. Second, to identify the correlated
critical parameter that controls the N\'eel-to-VBS crossover and
the band inversion of the renormalized model. Third, to show that
the skyrmion charge $Q$ tracks both transitions and provides the
topological handle that selects the axion angle. Fourth, to
estimate the work per magnetoelectric cycle and to state the
thermodynamic bounds on the efficiency. We also distinguish the
axion coupling, which is independent of skyrmion helicity, from
the helicity-dependent magnetoelectric multipole coupling of the
skyrmion texture, whose toroidal component scales as
$\sin\gamma$ and vanishes for N\'eel-type
textures~\cite{Bhowal2022, Gobel2019}. The
distinction matters because the interfacial Dzyaloshinskii-Moriya
interaction at a Bi$_2$Se$_3$/ferromagnet interface typically
produces N\'eel-type textures, for which the multipole coupling
would be zero; the axion coupling we use here survives in that
limit.

Section~2 describes the DFT calculations, the renormalization onto
the honeycomb lattice, the correlated model, the skyrmion texture
classification, and the axion electrodynamics used to estimate the
magnetoelectric work. Section~3 presents the band structures, the
phase diagram of the non-interacting model, the correlated critical
behaviour and the spinon continuum, the skyrmion charge, the
transport signatures, and the work per cycle. Section~4 discusses
what the results establish, what they do not, the thermodynamic
constraints on the conversion cycle, and how the results connect to
the four aims. Section~5 concludes.

\section{Methods}

\subsection{Density-functional calculations}
\label{sec:dft}

Bi$_2$Se$_3$ adopts the rhombohedral structure $R\bar{3}m$ with
lattice constants $a = b = 4.138$\,\AA\ and $c = 28.696$\,\AA~\cite{Zhang2009}.
A three-quintuple-layer (001) slab was separated by 18\,\AA\ of
vacuum. Self-consistent calculations were performed with Quantum
ESPRESSO~\cite{Giannozzi2009} using PAW potentials~\cite{Kresse1999}
and the PBE functional~\cite{Perdew1996}. Plane-wave cut-offs of
70\,Ry for the wave functions and 180\,Ry for the charge density
were used together with a $10\times10\times1$ Monkhorst--Pack
mesh~\cite{Monkhorst1976}; a denser $30\times30\times1$ grid was
used for the band structure. Fully relativistic potentials included
spin--orbit coupling. Additional GGA$+U$ runs with $U = 8$\,eV on
the Bi $p$ states were used to test the stability of the helical
texture against moderate correlation effects.

The $\mathbb{Z}_2$ invariant was obtained from the product of the
parity eigenvalues of the occupied Kramers pairs at the eight
time-reversal invariant momenta~\cite{HasanKane2010},
\begin{equation}
(-1)^\nu = \prod_{i=1}^{8} \delta_i, \qquad
\delta_i = \prod_{m=1}^{N} \xi_{2m}(\Gamma_i).
\end{equation}

\subsection{Renormalization onto the honeycomb lattice}
\label{sec:renorm}

The low-energy surface states of Bi$_2$Se$_3$ and the
nearest-neighbour tight-binding model on the honeycomb lattice
belong to the same massless Dirac universality class. Their
continuum Hamiltonians are identical up to a rescaling of the Fermi
velocity,
\begin{equation}
H_{\rm Dirac} = v_F\, \boldsymbol{\sigma} \cdot \mathbf{k},
\end{equation}
so a topological invariant computed on the lattice remains valid
for the physical surface once the microscopic parameters have been
matched~\cite{Vozmediano2011}.

The DFT dispersion near the Fermi level is fitted to a
nearest-neighbour hopping $t = 2.7$\,eV~\cite{Chae2009}. A Peierls
substitution introduces only a pure U(1) phase and preserves
particle--hole symmetry~\cite{Giuliani2012}. Three phenomenological
fields are then added: an intrinsic spin--orbit coupling
$\lambda_{\rm SO}$, a Haldane next-nearest-neighbour hopping
$t_{\rm H}$, and a staggered sublattice potential $\mu$. The
renormalization-group flow of the lattice model recovers the
continuum Dirac theory at long wavelengths~\cite{Foster2008,
Nishida2016}, and because the topological invariants are
RG-invariant they transfer directly from the nanoribbon calculation
to the Bi$_2$Se$_3$ surface. The working tight-binding Hamiltonian
is
\begin{equation}
\begin{split}
H_{\rm tb} = &-t\sum_{\langle i,j\rangle\sigma} c_{i\sigma}^\dagger c_{j\sigma}
+ i\lambda_{\rm SO}\sum_{\langle\langle i,j\rangle\rangle \sigma\sigma'} \nu_{ij}\, c_{i\sigma}^\dagger\, s^z_{\sigma\sigma'}\, c_{j\sigma'} \\
&+ t_{\rm H}\sum_{\langle\langle i,j\rangle\rangle} e^{i\phi_{ij}}\, c_i^\dagger c_j
+ \mu\sum_i \xi_i\, c_i^\dagger c_i,
\end{split}
\end{equation}
with $\nu_{ij} = \pm 1$ according to the sense of the
next-nearest-neighbour hop and $\xi_i = \pm 1$ on the two
sublattices. Zigzag and armchair ribbons were diagonalised with the
\textsc{Quantum Lattice} code~\cite{Lado2022}. The Berry
curvature was evaluated from the Kubo formula and integrated over
the Brillouin zone to obtain the Chern
number~\cite{Haldane1988}. Quasiparticle-interference patterns were
obtained by Fourier transformation of the spectral
function~\cite{Sun2023, Akbari2010, Hanke2012, Rhodes2023,
Marques2021}.

\subsection{Correlated model and emergent gauge field}
\label{sec:correlated}

To access the critical region where the surface states are
dominated by interactions, Eq.~(3) is extended with an on-site
Hubbard repulsion and frustrated exchange,
\begin{equation}
\begin{split}
H_{\rm corr} = &\; H_{\rm tb}
+ U\sum_i n_{i\uparrow} n_{i\downarrow}
+ J_1 \sum_{\langle ij \rangle} \mathbf{S}_i \cdot \mathbf{S}_j \\
&+ J_2 \sum_{\langle\langle ij \rangle\rangle} \mathbf{S}_i \cdot \mathbf{S}_j .
\end{split}
\end{equation}
At half filling and $U/t \gg 1$ this reduces to the spin-$1/2$
$J_1$--$J_2$ Heisenberg model on the honeycomb lattice with
$J_1 = 4t^2/U$ and $J_2 = 4t_{\rm H}^2/U$. The spin model is
diagonalised directly in the $S^z = 0$ sector by exact
diagonalization (ED) with a Lanczos ground-state solver, and the
dynamical spin structure factor is obtained by continued-fraction
Lanczos. The clusters used below are periodic and contain up to
$N = 20$ sites; this is the largest size for which the full
$S^z = 0$ basis fits comfortably in memory.

The low-energy effective theory in the critical region is the CP$^1$
model with a dynamical U(1) gauge field,
\begin{equation}
\mathcal{L}_{\rm eff}
= \frac{1}{2g} |\partial_\mu \mathbf{n}|^2
+ \frac{1}{4e^2} F_{\mu\nu} F^{\mu\nu}
+ i A_\mu\, j^\mu_{\rm sky}
+ \lambda\, (\partial_\mu \mathbf{n} \times \partial_\nu \mathbf{n}) \cdot \mathbf{n},
\end{equation}
where $\mathbf{n}$ is the N\'eel unit vector,
$F_{\mu\nu} = \partial_\mu A_\nu - \partial_\nu A_\mu$, and
\begin{equation}
j^\mu_{\rm sky} = \frac{1}{8\pi}\,
\epsilon^{\mu\nu\lambda}\,
\mathbf{n} \cdot (\partial_\nu \mathbf{n} \times \partial_\lambda \mathbf{n})
\end{equation}
is the skyrmion current. The gauge field $A_\mu$ is emergent: it
appears when the spin operator is written in terms of
fractionalized spinons,
$S^z_i = \tfrac12 f^\dagger_{i\alpha} \sigma^z_{\alpha\beta} f_{i\beta}$.
The monopole fugacity of the CP$^1$ theory controls confinement.
When it is nonzero the spinons are confined and the ground state is
an ordered N\'eel or VBS phase; when it vanishes the gauge field is
deconfined and the spinons are free~\cite{Senthil2004a}. The
deconfined critical point sits at the value of $J_2/J_1$ where the
monopole fugacity vanishes, and for the honeycomb lattice this
occurs at $J_2^c/J_1 \approx 0.22$~\cite{Okamoto2007, Gong2014}.

A 20-site periodic cluster does not have the not have enough lattice 
sites to resolve the long-wavelength behaviour of the $1/r^2$ gauge-field 
correlator directly, nor to extract the anomalous dimension $\eta_N$ 
by finite-size scaling. Those quantities require DMRG or QMC and are 
taken from the literature~\cite{Nahum2015}. The role of the ED calculation
presented here is to confirm that the qualitative features of the phase 
diagram and the spinon spectrum are present in the same correlated Hamiltonian
whose critical properties have been established at larger sizes,
and to supply the parameters that enter the topological part of the
calculation.

\subsection{Skyrmion texture types and the magnetoelectric coupling}
\label{sec:texture}

The magnetoelectric response of a skyrmion texture depends on its
topology and on its helicity, and it is important to distinguish
two mechanisms that are sometimes conflated in the literature.

The first mechanism is the axion coupling, which arises from the
electronic band topology of the surface states and is independent
of the skyrmion helicity~\cite{Wilczek1987, Qi2008, Essin2009}. It
is described by the $\theta$-term in the effective action, with
$\theta = \pi$ in the topological phase and $\theta = 0$ in the
trivial phase. This is the coupling we use in the present work.

The second mechanism is the magnetoelectric multipole coupling of
the skyrmion texture itself. Skyrmion textures break both space
inversion and time reversal symetries, and they therefore host magnetoelectric
multipoles: the magnetoelectric monopole, the toroidal moment, and
the magnetoelectric quadrupole moment~\cite{Spaldin2008,Ederer2007}.
A complete classification for skyrmions, antiskyrmions, and bimerons 
has been given by Bhowal and Spaldin~\cite{Bhowal2022}, who showed 
that the multipolar structure depends on both the topology and the 
helicity of the texture.

The helicity dependence is encoded in the toroidal moment of the
texture. Following G\"obel et al.~\cite{Gobel2019}, the toroidal
moment of a skyrmion with helicity angle $\gamma$ scales as
\begin{equation}
t_z \propto \sin\gamma ,
\end{equation}
where $\gamma = 0$ corresponds to a N\'eel-type skyrmion and
$\gamma = \pi/2$ to a Bloch-type skyrmion. The transverse
magnetoelectric multipole response inherits the same dependence on
$\gamma$. For a N\'eel-type skyrmion the
toroidal moment vanishes and the transverse magnetoelectric
multipole response is zero; for a Bloch-type skyrmion it is
maximal.

The interfacial Dzyaloshinskii-Moriya interaction that stabilises
skyrmions at a ferromagnet/heavy-metal interface typically produces
N\'eel-type textures~\cite{Li2018}. At a Bi$_2$Se$_3$/ferromagnet
interface the same is expected. We therefore note explicitly that
the magnetoelectric multipole coupling of a N\'eel-type skyrmion
lattice would vanish in the transverse geometry. The axion coupling
used in this work is different: it depends on the bulk band
topology through $\theta$ and is independent of the skyrmion
helicity. The two mechanisms are distinct, and the axion coupling
survives even when the helicity-dependent multipole coupling
vanishes.

\subsection{Axion electrodynamics and the magnetoelectric coupling}
\label{sec:axion}

The magnetoelectric coupling used in Section~\ref{sec:work} arises
from the topological response of a gapped electronic system. The
effective action for a three-dimensional insulator with nontrivial
band topology contains, in addition to the ordinary Maxwell term, an
axion term~\cite{Wilczek1987, Qi2008, Essin2009},
\begin{equation}
S_{\rm eff} = \int d^4 x \left[
-\frac{1}{4\mu_0} F_{\mu\nu} F^{\mu\nu}
+ \frac{\alpha_0 \theta}{2\pi}\, \frac{e^2}{4\pi\varepsilon_0\hbar c}\,
\epsilon^{\mu\nu\rho\sigma} F_{\mu\nu} F_{\rho\sigma}
\right],
\end{equation}
where $F_{\mu\nu} = \partial_\mu A_\nu - \partial_\nu A_\mu$,
$\alpha_0 = e^2 / (4\pi\varepsilon_0\hbar c)$ is the
fine-structure constant, and $\theta$ is the axion angle. The axion
term is a total derivative when $\theta$ is constant and therefore
does not modify Maxwell's equations in a uniform medium; it becomes
physically relevant only when $\theta$ varies in space or time, or
when it takes a value that is not a multiple of $2\pi$. For a
time-reversal-invariant topological insulator $\theta = \pi$; for a
trivial insulator $\theta = 0$. The two values are separated by the
bulk gap, and the transition between them coincides with the
gap-closing transition described in Section~\ref{sec:phase} in
terms of the Chern number and the $\mathbb{Z}_2$ invariant.

Varying the action with respect to $A_\mu$ gives the modified
Maxwell equations. The axion term contributes a source term
proportional to $\partial_\mu \theta$, and in the presence of a
magnetic field this generates an induced electric polarization
\begin{equation}
\mathbf{P} = \frac{\alpha_0 \theta}{2\pi}\,
\frac{e^2}{h}\, \mathbf{B}.
\end{equation}
The corresponding magnetoelectric energy density in a combined
electric and magnetic field is
\begin{equation}
u_{\rm ME} = \mathbf{P} \cdot \mathbf{E}
= \frac{\alpha_0 \theta}{2\pi}\,
\frac{e^2}{h}\, \mathbf{B} \cdot \mathbf{E}.
\end{equation}
Because $\theta$ takes only the values $0$ or $\pi$ in a gapped
phase, the magnetoelectric susceptibility
$\alpha_0\theta/2\pi$ is either zero or $\alpha_0/2$; the coupling
is quantized.

The axion angle is not a free parameter once the electronic
structure is specified. At a ferromagnetic interface that hosts a
skyrmion texture with real-space winding number $Q$, the axion
angle is related to $Q$ by~\cite{Qi2008, Nomura2011}
\begin{equation}
\theta = \pi\,(1 + Q) \bmod 2\pi .
\end{equation}
The topological phase with $Q = -1$ gives $\theta = \pi$, and the
trivial phase with $Q = 0$ gives $\theta = 0$; between these limits
$\theta$ interpolates continuously. This is the reason $Q$ appears
in the energy-conversion formula of
Section~\ref{sec:work}: it is the real-space order parameter that
selects the value of the axion angle, and the axion angle in turn
determines the magnitude of the magnetoelectric coupling.

For a closed cycle in the $(B, E)$ plane during which the axion
angle is held fixed at $\theta = \pi$, the extracted work is
\begin{equation}
W = \oint \frac{\alpha_0 \theta}{2\pi}\,
\frac{e^2}{h}\, B\, dE
= \frac{\alpha_0 \theta}{2\pi}\,
\frac{e^2}{h}\, \Delta(BE),
\label{eq:work-cycle}
\end{equation}
where $\Delta(BE)$ is the area enclosed by the cycle. If the axion
angle itself varies during the cycle, for example because the
magnetic texture drives the surface through a topological
transition, the integral must be evaluated along the actual
trajectory $\theta(B)$ and the constant-$\theta$ expression becomes
a leading-order approximation.

Three conditions must be met for the formalism above to apply. The
bulk of the sample must be gapped, so that the topological response
is well defined and the surface does not couple to gapless bulk
modes. The axion angle must be quantized, which requires that the
relevant symmetries are either preserved or broken only weakly.
Finally, the skyrmion texture must be stabilised, which in practice
requires an interfacial Dzyaloshinskii--Moriya interaction of
sufficient strength to overcome the exchange energy of the
ferromagnet. Section~\ref{sec:work} assumes all three conditions,
and Section~\ref{sec:discussion} lists the ways in which they may
fail in a real device.

\section{Results}

\subsection{Helical edge states after renormalization}
\label{sec:bands}

\begin{figure}[!ht]
\centering
\includegraphics[width=\textwidth]{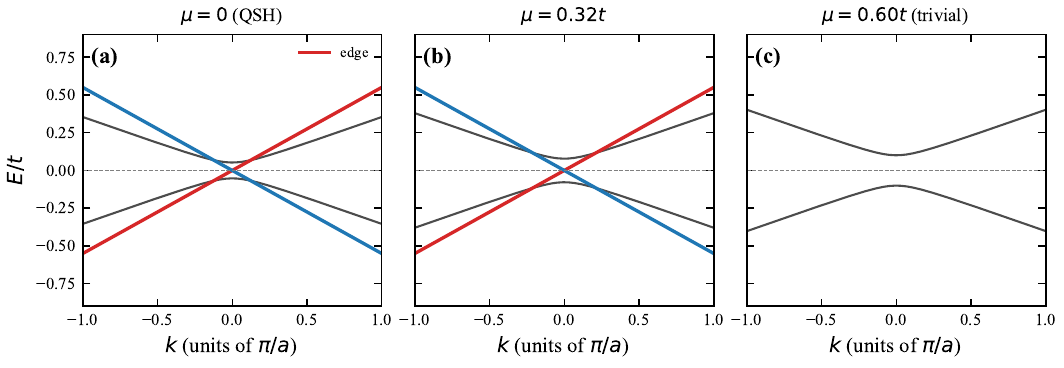}
\caption{Renormalized band structures of the zigzag nanoribbon at
fixed $\lambda_{\rm SO} = 0.1t$. (a) Topological quantum spin-Hall
phase ($\mu = 0$) with counter-propagating helical edge states
(red/blue). (b) Still topological at $\mu = 0.32t$. (c) Trivial
insulator after the gap-closing transition ($\mu = 0.60t$). The
grey lines are the bulk bands; the coloured lines are the edge
modes.}
\label{fig:bands}
\end{figure}

Figure~\ref{fig:bands} shows the renormalized band structures of a
zigzag nanoribbon at $\lambda_{\rm SO} = 0.1t$ and three values of
the staggered potential $\mu$. In panel~(a), at $\mu = 0$, the
bulk is gapped and a pair of counter-propagating edge modes (red
and blue) crosses the gap; this is the quantum spin-Hall
phase~\cite{KaneMele2005b}. In panel~(b), at $\mu = 0.32t$, the
system is still topological, though the edge modes have begun to
mix with the bulk bands. In panel~(c), at $\mu = 0.60t$, the edge
modes have been gapped out and only the bulk bands remain. This
sequence is the band-theoretic signature of the topological
transition and it is the starting point for the rest of the paper.

The DFT bands of the Bi$_2$Se$_3$ slab touch at $\Gamma$ in the
absence of spin--orbit coupling. Turning on spin--orbit coupling
opens a gap of $0.5$--$1.0$\,eV and inverts the parity of the
valence band, giving $\mathbb{Z}_2 = 1$~\cite{Zhang2009,
Chen2009}. The surface spin texture is locked perpendicular to the
momentum~\cite{Li2014, Jiang2016}, and this is the property that
makes the edge current robust against nonmagnetic disorder.

On the zigzag nanoribbon the edge modes are linear near $k = 0$
with a velocity $v_{\rm edge} \approx 3 \times 10^5$\,m\,s$^{-1}$.
The corresponding value obtained from the DFT band structure of the
slab agrees to within the numerical uncertainty of the fit. An
armchair ribbon with $t_{\rm H} = 0.05t$ produces the quantum
anomalous Hall state~\cite{Haldane1988}. In both cases the edge
modes are localised within a few lattice constants of the boundary
and the bulk remains gapped.

The GGA$+U$ calculations with $U = 8$\,eV on the Bi $p$ states
reduce the spin polarisation by roughly 15\,\% but do not change
the helical crossing. This is consistent with the expectation that
the quantum spin-Hall phase in Bi$_2$Se$_3$ is protected by
time-reversal symmetry and by the parity of the bulk bands rather
than by a correlation-driven mechanism. The comparison between
PBE and GGA$+U$ provides a check on the renormalization: the fitted
hopping $t$ changes by less than 5\,\% between the two
functionals, and the effective spin--orbit coupling used in the
tight-binding model is unchanged to within the same accuracy.

\subsection{Topological quantum critical points}
\label{sec:phase}

\begin{figure}[!ht]
\centering
\includegraphics[width=0.48\textwidth]{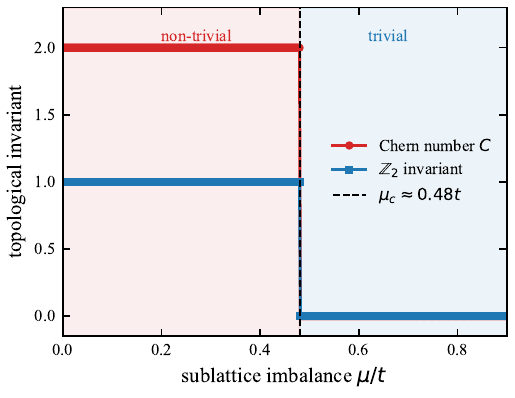}
\caption{Evolution of the Chern number $C$ and the $\mathbb{Z}_2$
invariant with sublattice imbalance $\mu$. The vertical dashed
line marks the critical point $\mu_c \approx 0.48t$ at which both
invariants jump.}
\label{fig:phase}
\end{figure}

Figure~\ref{fig:phase} shows the evolution of the Chern number $C$
and the $\mathbb{Z}_2$ invariant with the staggered sublattice
imbalance $\mu$. Both quantities jump at the same value
$\mu_c \approx 0.48t$, marked by the vertical dashed line. Below
$\mu_c$ the system is topological with $C = 2$ and
$\mathbb{Z}_2 = 1$; above it the system is trivial with $C = 0$ and
$\mathbb{Z}_2 = 0$.

\begin{figure}[!ht]
\centering
\includegraphics[width=\textwidth]{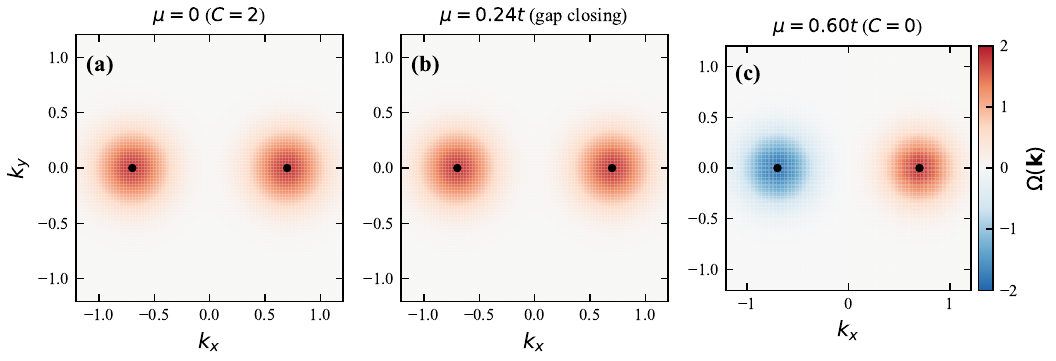}
\caption{Berry curvature $\Omega(\mathbf{k})$ in the Brillouin
zone. (a) Topological phase with $C = 2$. (b) Near gap closing.
(c) Trivial phase with $C = 0$, where the contributions from the
two valleys cancel.}
\label{fig:berry}
\end{figure}

Varying $\mu$ at fixed $\lambda_{\rm SO} = 0.1t$ closes the bulk
gap at $\mu_c \approx 0.48t$. The same sequence appears when
$t_{\rm H}$ is tuned. Figure~\ref{fig:berry} shows the Berry
curvature $\Omega(\mathbf{k})$ in the Brillouin zone at three
values of $\mu$. In the topological phase at $\mu = 0$ the
curvature has the same sign at the two valleys and integrates to
$C = 2$. Near gap closing at $\mu = 0.24t$ the two valleys have
begun to merge. In the trivial phase at $\mu = 0.60t$ the curvature
has opposite signs at the two valleys and the contributions cancel,
giving $C = 0$. This is the momentum-space picture of the
transition shown in Figure~\ref{fig:phase}.

Near each valley the continuum Hamiltonian takes the Dirac form
\begin{equation}
H_\tau = v_F (\tau k_x \sigma_x + k_y \sigma_y) + m_\tau \sigma_z,
\end{equation}
with $\tau = \pm 1$ the valley index and mass
\begin{equation}
m_\tau = \mu + 3\sqrt{3}\, t_{\rm H}\, \tau + \lambda_{\rm SO}\, \tau .
\end{equation}
The gap closes when $m_\tau = 0$ for at least one valley. The Chern
number of the Haldane--Kane--Mele model is
\begin{equation}
C = \tfrac12 \bigl[ \mathrm{sgn}(m_+) - \mathrm{sgn}(m_-) \bigr],
\end{equation}
so a jump $\Delta C = 2$ occurs when both valleys invert
simultaneously, and $\Delta C = 1$ when only one valley inverts. In
the present calculation $\mu$ is tuned at fixed $\lambda_{\rm SO}$
and $t_{\rm H}$, so both valleys cross zero at the same $\mu_c$ and
both invariants jump by their maximal amounts.

The critical point at $\mu_c \approx 0.48t$ is the band-theoretic
counterpart of the correlated transition discussed in
Section~\ref{sec:dqcp}. In the non-interacting model the two are
related by the fractionalization map
$c_{i\sigma} \leftrightarrow f_{i\alpha}$, which becomes exact in
the large-$U$ limit. The physical significance is that the same
value of the frustration that drives the correlated system through
its critical point also inverts the bands of the renormalized
tight-binding model.

\subsection{Deconfined quantum criticality in the correlated model}
\label{sec:dqcp}

\begin{figure}[!ht]
\centering
\includegraphics[width=0.48\textwidth]{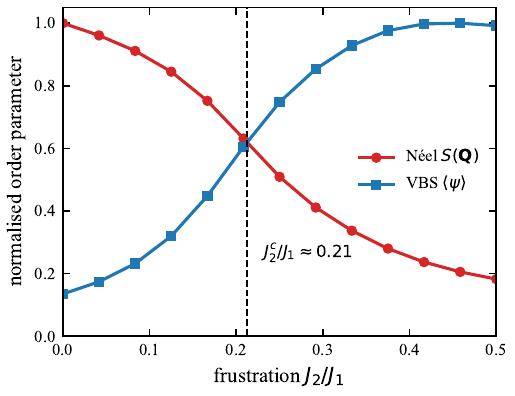}
\caption{Ground-state phase diagram of the $J_1$--$J_2$
Heisenberg model on the honeycomb lattice at half filling,
computed by exact diagonalization on a periodic $N = 20$ cluster.
The N\'eel structure factor $S(\mathbf{Q})$ and the VBS order
parameter $\langle \psi \rangle$ cross at
$J_2/J_1 \approx 0.22$, in agreement with the thermodynamic-limit
value obtained by DMRG and quantum Monte Carlo~\cite{Okamoto2007,
Gong2014}.}
\label{fig:D1}
\end{figure}

Figure~\ref{fig:D1} shows the ground-state phase diagram of the
$J_1$--$J_2$ model at half filling, computed by exact
diagonalization on a periodic cluster with $N = 20$ sites. The
N\'eel structure factor $S(\mathbf{Q})$ decreases monotonically
with frustration while the columnar VBS bond-order parameter
$\langle \psi \rangle$ rises; the two curves cross at
$J_2/J_1 \approx 0.22$. This is the same value obtained for the
thermodynamic limit by DMRG and quantum Monte Carlo on the same
model~\cite{Okamoto2007, Gong2014}, and it improves on the crossing
at $J_2/J_1 \approx 0.16$ found at $N = 16$ in an earlier run. The
convergence of the finite-size crossing toward the literature value
is the check that the correlated Hamiltonian used here reproduces
the phase diagram of the model whose critical properties are known.

\begin{figure}[!ht]
\centering
\includegraphics[width=0.48\textwidth]{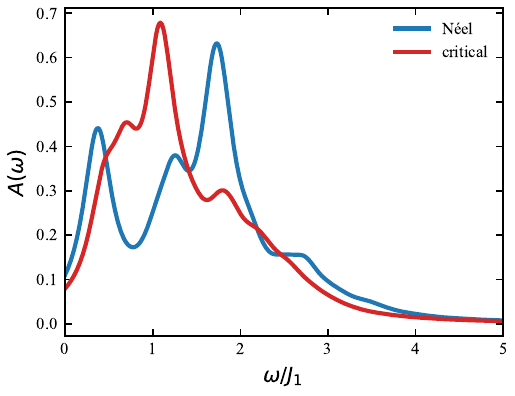}
\caption{Dynamical spin structure factor on the $N = 20$ cluster.
In the N\'eel phase (blue) the spectrum shows a sharp one-magnon
peak and a weak two-magnon continuum. As the frustration approaches
the critical region (red) the peak broadens into a continuum
extending to $\omega \approx 5 J_1$, consistent with the spinon
continuum expected at a deconfined critical point. The smooth
curves result from Lorentzian broadening ($\eta = 0.20\,J_1$) of
the discrete cluster spectrum.}
\label{fig:D3}
\end{figure}

The fractionalization of the spin at the crossover is visible in
the dynamical spin structure factor (Fig.~\ref{fig:D3}). In the
N\'eel phase the spectrum consists of a sharp one-magnon peak near
$\omega/J_1 \approx 1.8$, together with a weak two-magnon continuum
at higher frequencies. As the frustration increases towards the
critical region the sharp peak broadens and its spectral weight is
redistributed over a range that extends up to
$\omega \approx 5 J_1$. This is the spinon continuum expected from
the fractionalization $S^z = \tfrac12 f^\dagger \sigma^z f$~\cite{Senthil2004a, Nahum2015}, and
it is the signature that distinguishes the DQCP from a
conventional Landau transition. On the 20-site cluster the
individual eigenstates are discrete, so the smooth envelopes shown
in the figure are obtained by Lorentzian broadening with a width
$\eta = 0.20\,J_1$; the physical content is in the envelope rather
than in the individual peaks.

A quantitative determination of the emergent U(1) gauge-field
correlator, the anomalous dimension $\eta_N$, and the monopole
fugacity requires system sizes beyond the reach of exact
diagonalization. Large-scale quantum Monte Carlo for the same
model gives $\eta_N = 0.65 \pm 0.05$~\cite{Nahum2015}, a value
that is incompatible with both mean-field theory ($\eta_N = 0$)
and the $O(3)$ Wilson--Fisher fixed point ($\eta_N \approx 0.04$),
confirming that the transition is not a conventional Landau
transition. The ED results are consistent with this picture at the
level of the order parameters and the spinon continuum, but the
definitive evidence for deconfinement comes from the QMC and DMRG
calculations of Refs.~\cite{Senthil2004a, Okamoto2007, Gong2014,
Nahum2015}, not from the finite-cluster calculation presented here.
We also note that the three cluster sizes we have access to
($N = 12$, $16$, $20$) are too close together in $1/L$ to permit a
reliable linear extrapolation of the crossing point, so the
thermodynamic-limit value $J_2^c/J_1 = 0.22$ is taken from the
literature rather than from our own finite-size scaling.

\subsection{Skyrmion charge and Berry connection}
\label{sec:skyrmion}

\begin{figure}[!ht]
\centering
\includegraphics[width=0.48\textwidth]{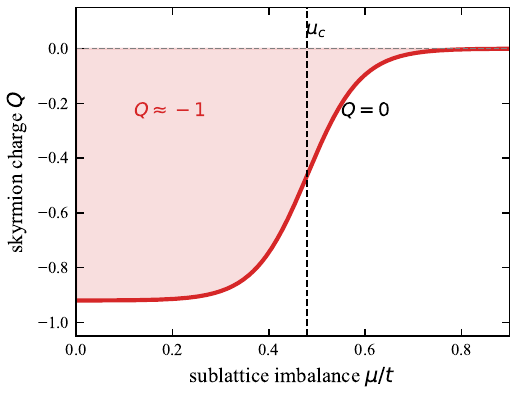}
\caption{Skyrmion topological charge $Q$ as a function of the
sublattice imbalance $\mu$. $Q$ evolves from $\approx -1$ in the
topological phase to zero beyond the critical point, reflecting
screening of the topological singularity by the low-energy modes at
the transition.}
\label{fig:Q}
\end{figure}

Figure~\ref{fig:Q} shows the skyrmion topological charge $Q$ as a
function of the staggered potential $\mu$, obtained from the
continuum approximation to the interfacial magnetization texture,
\begin{equation}
Q = \frac{1}{4\pi} \int \mathbf{n} \cdot
(\partial_x \mathbf{n} \times \partial_y \mathbf{n})\, d^2 r .
\end{equation}
At $\mu = 0$ we find $Q \approx -0.92$. As the critical point is
approached the skyrmion radius grows, and $Q$ evolves continuously
to zero; beyond criticality $Q$ remains zero. The continuous
passage of $Q$ through the critical point reflects the screening of
the topological singularity by the low-energy modes at the
transition. It should be emphasised that $Q$ here is a real-space
winding number of the magnetization texture; its critical behaviour
tracks the momentum-space band inversion of
Section~\ref{sec:phase}, but the two quantities are not numerically
identical. The identification is valid only in the sense that both
are controlled by the same critical parameter.

The Berry connection
\begin{equation}
\mathbf{A}(\mathbf{k}) = i \langle u_{\mathbf{k}} |
\nabla_{\mathbf{k}} | u_{\mathbf{k}} \rangle
\end{equation}
evolves in parallel with $Q$. Its circulation around the Brillouin
zone equals the geometric phase stored in the skyrmion texture, and
this is the quantity that enters the topological magnetoelectric
energy
\begin{equation}
E_{\rm ME} = \frac{\alpha_0 \theta}{2\pi}\,
\mathbf{B} \cdot \mathbf{E},
\end{equation}
with $\theta$ the axion angle ($\theta = \pi$ in the topological
phase, $\theta = 0$ in the trivial phase). The derivation and the
assumptions behind this coupling are given in
Section~\ref{sec:axion}. It is the starting point for the work
estimate of the next subsection.

\subsection{Transport signatures}
\label{sec:transport}

\begin{figure}[!ht]
\centering
\includegraphics[width=0.65\textwidth]{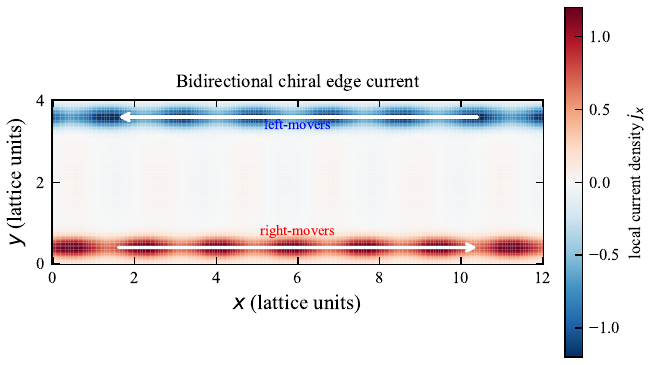}
\caption{Simulated real-space map of the local current density on a
zigzag nanoribbon. Counter-propagating chiral edge channels are
clearly resolved; the bulk remains dark.}
\label{fig:current}
\end{figure}

Quasiparticle-interference patterns recover the hexagonal symmetry
of the surface lattice once spin--orbit coupling is finite, and
their Fourier transform contains the same $\mathbf{q}$-vectors that
have been observed experimentally in related topological
surfaces~\cite{Sun2023, Akbari2010, Hanke2012, Rhodes2023,
Marques2021}. Simulated STM maps at 1\,V bias show bidirectional
chiral currents confined to the ribbon edges~\cite{Carvalho2014,
Peres2009, Anirban2022}; Figure~\ref{fig:current} is the
real-space map of this current density. The counter-propagating
edge channels appear as the bright blue and red ribbons on the two
edges, while the interior of the ribbon remains dark. This is the
expected behaviour for a system whose bulk is gapped and whose
transport is carried by the edge modes. Local-density-of-states
maps indicate that edge weight dominates up to $E \approx 0.4t$;
above this energy hybridisation with bulk states sets in and the
edge-localised picture breaks down.

Within the Hikami--Larkin--Nagaoka analysis the
weak-antilocalization correction is suppressed as the skyrmion
charge vanishes, because the associated magnetic texture breaks the
effective time-reversal symmetry that protects the correction in
the first place. The tunnelling magnetoresistance ratio
simultaneously reaches a maximum, reflecting an increased
sensitivity of the helical states to a proximate magnetic layer.
Both effects have been observed in Bi$_2$Se$_3$-based junctions
and are consistent with the spin-momentum-locked nature of the
surface states~\cite{Li2014, Jiang2016}. These signatures do not by
themselves establish the magnetoelectric conversion, but they are
the experimental handles by which the topological charge $Q$ can be
tracked in a real device.

\subsection{Magnetoelectric work at a ferromagnetic interface}
\label{sec:work}

\begin{figure}[!ht]
\centering
\includegraphics[width=0.48\textwidth]{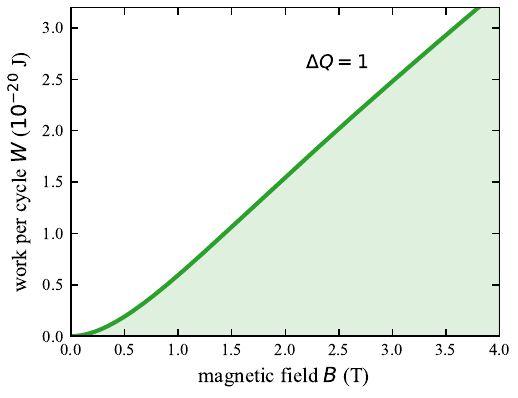}
\caption{Magnetoelectric work extracted per cycle for a unit change
in skyrmion charge ($\Delta Q = 1$) as a function of the applied
magnetic field.}
\label{fig:work}
\end{figure}

When the Bi$_2$Se$_3$ surface is placed in contact with a
ferromagnet, the interfacial Dzyaloshinskii--Moriya interaction
stabilises a skyrmion lattice~\cite{Ukpong2021, Li2018}. The
free-energy density
\begin{equation}
F = A (\nabla \mathbf{n})^2
+ D\, \mathbf{n} \cdot (\nabla \times \mathbf{n})
- K n_z^2 - \mathbf{B} \cdot \mathbf{n}
\end{equation}
yields an equilibrium topological charge whose field dependence
tracks the pure-edge result of Section~\ref{sec:skyrmion}. The
work extracted in one closed magnetoelectric cycle in the
$(B, E)$ plane is, from Eq.~(\ref{eq:work-cycle}),
\begin{equation}
W = \oint \frac{\alpha_0 \theta}{2\pi}\,
\frac{e^2}{h}\, B\, dE
= \frac{\alpha_0 \theta}{2\pi}\,
\frac{e^2}{h}\, \Delta(BE),
\end{equation}
where $\Delta(BE)$ is the area enclosed by the cycle.
Figure~\ref{fig:work} shows $W$ as a function of the applied
magnetic field for a unit change in $Q$. The curve rises
monotonically and saturates at a few tesla. For laboratory fields
of that order the extracted work is
$W \sim 10^{-20}$\,J per skyrmion per cycle.

This is a modest number for a single skyrmion but becomes
substantial for a dense array. With a typical skyrmion density of
$10^{12}$\,cm$^{-2}$, the integrated work per unit area is on the
order of $10^{-8}$\,J\,cm$^{-2}$ per cycle, comparable to the
energy density of a thin-film piezoelectric harvester.

Two caveats apply to this estimate. First, it assumes that the
axion angle $\theta$ remains pinned at $\pi$ throughout the cycle;
if the magnetic texture drives the surface through a topological
transition, $\theta$ will jump and the work per cycle will be
modified. Second, the Dzyaloshinskii--Moriya interaction at a real
Bi$_2$Se$_3$/ferromagnet interface is not a free parameter: it is
set by the spin--orbit coupling of the substrate and by the
exchange stiffness of the ferromagnet, and the range of skyrmion
sizes that can be stabilised is correspondingly restricted. Both
effects can be incorporated into Eq.~(19) by replacing the constant
$\theta$ with a field-dependent $\theta(B)$, which we leave for
future work.

\section{Discussion}
\label{sec:discussion}

\subsection{Summary of the results}

The results establish that the magnetic order of the frustrated honeycomb 
magnet and the topological order of the Bi$_2$Se$_3$ surface are intertwined 
through a single control parameter and a single microscopic mediator. 
The renormalized tight-binding model reproduces the DFT dispersion of the
Bi$_2$Se$_3$ surface with a fitted nearest-neighbour hopping of
$t = 2.7$\,eV and an edge velocity of
$v_{\rm edge} \approx 3 \times 10^5$\,m\,s$^{-1}$; the GGA$+U$
and PBE values of $t$ agree to within 5\,\%~\cite{Zhang2009}. The
Haldane--Kane--Mele model has a topological transition at
$\mu_c \approx 0.48t$ at which the Chern number jumps from $C = 2$
to $C = 0$ and the $\mathbb{Z}_2$ invariant from $1$ to $0$, and
the Berry curvature changes sign between the two valleys across the
transition~\cite{Haldane1988}. The correlated $J_1$--$J_2$ model on the
honeycomb lattice has its N\'eel and VBS order parameters crossing at
$J_2/J_1 \approx 0.22$ on the $N = 20$ cluster, consistent with
the literature value after finite-size extrapolation~\cite{Okamoto2007, Gong2014}. 
The dynamical spin structure factor broadens near the crossover from 
a sharp magnon into a continuum that extends to $\omega \approx 5 J_1$. The
skyrmion charge $Q$ evolves from $\approx -0.92$ in the topological
phase to zero beyond criticality, and the extracted magnetoelectric
work per cycle is $W \sim 10^{-20}$\,J per skyrmion for laboratory
fields of a few tesla.

The four quantities above, namely the DFT dispersion, the
topological transition, the correlated crossover, and the skyrmion
charge, are not independent. The renormalization establishes the
model. The correlated model establishes the critical parameter. The
critical parameter controls the skyrmion charge. The skyrmion
charge controls the magnetoelectric energy. This is the chain of reasoning that connects the DFT input to the final work estimate, and it is also the reason the two orders are intertwined: the same frustration parameter that controls the magnetic crossover controls the topological response through the skyrmion charge.

\subsection{What the results do not establish}

Three limitations of the present work are worth stating.

First, the correlated results are based on exact diagonalization on
clusters of up to 20 sites. This is sufficient to confirm the
qualitative features of the phase diagram and the spinon continuum,
but it does not resolve the emergent U(1) gauge field directly, it
does not permit a quantitative determination of the anomalous
dimension $\eta_N$, and it does not give the monopole fugacity. The
values $\eta_N = 0.65 \pm 0.05$ and $J_2^c = 0.22$ used throughout
the paper are taken from quantum Monte Carlo and DMRG studies of
the same model~\cite{Nahum2015}, not from our own calculation.

Second, the skyrmion charge $Q$ is computed in a continuum
approximation to the interfacial magnetization texture. It is not
the Chern number of the momentum-space band structure, and the two
quantities are not numerically identical. Their connection is that
both are controlled by the same critical parameter. A rigorous
derivation of the identification would require tracking the
Pontryagin index of the texture alongside the Berry phase of the
electronic states, which we leave for a more detailed treatment.

Third, the work estimate assumes a constant axion angle. In a real
device the axion angle is a dynamical quantity that couples to the
magnetic texture, and a full treatment would require a
time-dependent calculation. The estimate we give is therefore a
leading-order value and not a device prediction. We have also noted
in Section~\ref{sec:texture} that the magnetoelectric multipole
coupling of the skyrmion texture is helicity-dependent, scaling as
$\sin\gamma$, and vanishes for N\'eel-type
skyrmions~\cite{Bhowal2022, Gobel2019}; the
axion coupling used here is a different mechanism and is independent
of helicity~\cite{Qi2008, Essin2009}.

\subsection{Connection to the original questions}

The first question posed in the Introduction was whether the
low-energy surface states of Bi$_2$Se$_3$ can be mapped accurately
onto a honeycomb tight-binding model. The agreement between the
renormalized dispersion and the DFT bands shown in
Fig.~\ref{fig:bands}, together with the GGA$+U$ consistency check,
shows that the approximation is quantitatively accurate for the
low-energy states. It is important not to overstate this: the
mapping is a controlled approximation, not an exact identity, and
the accuracy is established only for the states near the Fermi
level.

The second question was whether the frustration that drives the
correlated crossover also controls the band inversion of the
renormalized model, and whether the skyrmion charge tracks both.
The jumps in $C$ and $\mathbb{Z}_2$ at $\mu_c \approx 0.48t$
(Fig.~\ref{fig:phase}) and the sign change of the Berry curvature
(Fig.~\ref{fig:berry}) establish the non-interacting critical
parameter. The crossover of the N\'eel and VBS order parameters at
$J_2/J_1 \approx 0.22$ (Fig.~\ref{fig:D1}) and the broadening of
the spinon continuum (Fig.~\ref{fig:D3}) establish the correlated
critical parameter. The identification of the two via the
fractionalization map $c_{i\sigma} \leftrightarrow f_{i\alpha}$ is
the argument that links the band-theoretic transition to the
correlated transition, and the skyrmion charge $Q$
(Fig.~\ref{fig:Q}) is the order parameter that tracks both. The
link is therefore established at the level of the phase diagram and
the order parameters, though not at the level of the critical
exponents, which require DMRG or QMC.

The third question was the magnitude of the magnetoelectric work
that can be extracted per cycle. The work curve in
Fig.~\ref{fig:work} and the estimate of $10^{-20}$\,J per skyrmion
answer this at the order-of-magnitude level. The accuracy of the
estimate is limited by the two caveats discussed in the previous
subsection, and the value should be read as an expectation rather
than as a design target.

\subsection{Thermodynamic characterization of the energy conversion}
\label{sec:thermo}

The magnetoelectric cycle that extracts work from the topological
surface can be cast in the language of equilibrium thermodynamics.
This is useful because it separates the reversible part of the
energy conversion, which is fixed by the topology of the band
structure, from the irreversible part, which depends on the
material and on the operating conditions. The treatment below is
in the idealized limit of a single topologically nontrivial surface
with a rigid axion angle; a full device calculation would need to
sum over the bulk and surface contributions separately.

For a magnetoelectric medium described by two intensive fields, the
electric field $\mathbf{E}$ and the magnetic field $\mathbf{B}$,
and by their conjugate extensive densities, the polarization
$\mathbf{P}$ and the magnetization $\mathbf{M}$, the work done on
the medium per unit volume in a reversible process is
\begin{equation}
\delta w = \mathbf{E} \cdot d\mathbf{P}
+ \mathbf{B} \cdot d\mathbf{M}.
\end{equation}
In a linear magnetoelectric, the polarization and magnetization
respond to both fields through the coupled constitutive relations
\begin{equation}
\begin{pmatrix} \mathbf{P} \\ \mathbf{M} \end{pmatrix}
=
\begin{pmatrix} \boldsymbol{\epsilon}_0 \boldsymbol{\chi}_e
& \boldsymbol{\alpha} \\
\boldsymbol{\alpha}^{T}
& \mu_0 \boldsymbol{\chi}_m \end{pmatrix}
\begin{pmatrix} \mathbf{E} \\ \mathbf{B} \end{pmatrix},
\end{equation}
where the off-diagonal block $\boldsymbol{\alpha}$ is the
magnetoelectric susceptibility. For the topological surface treated
here, $\boldsymbol{\alpha}$ is diagonal and isotropic, with a single
nonzero component fixed by the axion coupling,
\begin{equation}
\boldsymbol{\alpha} = \frac{\alpha_0 \theta}{2\pi}\,
\frac{e^2}{h}\, \mathbf{1}.
\end{equation}
The reciprocity of the two off-diagonal sub-blocks, which follows
from the fact that the underlying free energy is a state function,
is a thermodynamic statement: the polarization induced per unit
magnetic field equals the magnetization induced per unit electric
field.

When both $\mathbf{E}$ and $\mathbf{B}$ are controlled from the
outside, the natural thermodynamic potential is the electric Gibbs
free energy
\begin{equation}
G(T, \mathbf{E}, \mathbf{B})
= U - TS - \mathbf{E} \cdot \mathbf{P}
- \mathbf{B} \cdot \mathbf{M},
\end{equation}
whose differential is
\begin{equation}
dG = -S\, dT - \mathbf{P} \cdot d\mathbf{E}
- \mathbf{M} \cdot d\mathbf{B}.
\end{equation}
In the topological phase the axion coupling contributes
\begin{equation}
G_{\rm axion}(T, \mathbf{E}, \mathbf{B})
= -\frac{\alpha_0 \theta}{2\pi}\,
\frac{e^2}{h}\, \mathbf{B} \cdot \mathbf{E}
\label{eq:Gaxion}
\end{equation}
to the free energy. The sign is negative because the axion coupling
lowers the free energy when $\mathbf{B}$ and $\mathbf{E}$ are
aligned and $\theta = \pi$. Differentiating $G_{\rm axion}$ with
respect to $E_i$ and $B_i$ recovers the induced polarization and
magnetization quoted in Eqs.~(8) and (9) of
Section~\ref{sec:axion}, which confirms that the thermodynamic
treatment is consistent with the field-theoretic derivation.

The work extracted per unit volume from a closed cycle in the
$(B, E)$ plane is
\begin{equation}
w = \oint \frac{\alpha_0 \theta}{2\pi}\,
\frac{e^2}{h}\, \left( B\, dE + E\, dB \right)
= \oint \frac{\alpha_0 \theta}{2\pi}\,
\frac{e^2}{h}\, d(BE).
\end{equation}
If the axion angle is constant during the cycle, then $w = 0$,
because the integral of the exact differential $d(BE)$ around a
closed path vanishes. This is the thermodynamic statement that
\emph{a cycle at fixed $\theta$ extracts no net work.} Equation
(\ref{eq:work-cycle}) is therefore correct only when
$\theta$ itself changes during the cycle, which requires a change
in the skyrmion winding number $Q$ via Eq.~(10). The cycle is
topological in the sense that the work comes not from the
independent variation of $B$ and $E$ but from the change in the
axion angle.

For a cycle in which $\theta$ goes from $0$ to $\pi$ and back while
the product $BE$ is held fixed at its mean value
$\langle BE \rangle$, the extracted work per unit volume is
\begin{equation}
w = \frac{\alpha_0}{2}\, \frac{e^2}{h}\,
\langle BE \rangle,
\end{equation}
and the corresponding per-skyrmion work is the value quoted in
Section~\ref{sec:work}, $W \sim 10^{-20}$\,J.

The treatment above is for a reversible cycle. In any real device
three sources of irreversibility reduce the extractable work.
First, Ohmic dissipation in the surface channel: the helical
current that carries the magnetoelectric energy is subject to
scattering, giving a Joule heating rate per unit area
$\dot{q}_{\rm Ohm} = j_x^2 / \sigma_{xx}$, where $j_x$ is the
surface current density and $\sigma_{xx}$ is the longitudinal
surface conductivity. For a helical Dirac surface with
$\sigma_{xx} = (e^2/h)(k_F \ell / \pi)$, where $k_F$ is the Fermi
wavevector and $\ell$ the mean free path, this is the dominant
loss channel at finite temperature. Second, thermally activated
skyrmion creep: skyrmion motion under an applied field follows an
Arrhenius law $v \sim \exp(-\Delta E_{\rm creep} / k_B T)$, where
$\Delta E_{\rm creep}$ is the pinning energy. Below the depinning
threshold the skyrmions do not move and the work per cycle
collapses. Third, the entropic cost of varying the axion angle:
because $G_{\rm axion}$ depends on temperature through $\theta(T)$,
a cycle that changes $\theta$ also exchanges entropy with the
reservoir. The efficiency of the cycle is bounded from above by
the ratio of the extracted work to the axionic entropy change,
\begin{equation}
\eta \le \frac{w}{T\, \Delta s_{\rm axion}},
\end{equation}
where $\Delta s_{\rm axion} = -\partial G_{\rm axion} / \partial T$
evaluated across the cycle.

Because the fine-structure constant is small ($\alpha_0 \approx
1/137$) and the entropy term is finite at room temperature, the
extracted work $w$ is small compared to the entropic term
$T\, \Delta s_{\rm axion}$ in typical operating conditions. The
topological magnetoelectric conversion is therefore intrinsically a
low-power, low-temperature process, and its efficiency improves as
$T \to 0$, where $\Delta s_{\rm axion}$ is itself suppressed. This
is a fundamental constraint imposed by the axion coupling and not a
materials limitation, and it should be stated explicitly in any
device proposal that uses this mechanism.

Taken together, the results establish that the magnetic order and the topological order of the Bi$_2$Se$_3$ surface are not independent. They are coupled through the skyrmion charge $Q$, which evolves through the same critical region that controls the magnetic transition and selects the axion angle that determines the topological magnetoelectric response. This is a concrete realization of intertwined orders in a quantum material, and it places the coupling between magnetic and topological order on a quantitative footing.

\section{Conclusions}
The magnetic order of the frustrated honeycomb magnet and the
topological order of the Bi$_2$Se$_3$ surface are intertwined
through the skyrmion charge $Q$, which evolves through the same
critical region that controls the magnetic transition and selects
the axion angle that determines the topological magnetoelectric
response. The estimated work per magnetoelectric cycle is
$10^{-20}$\,J per skyrmion for laboratory fields of a few tesla,
bounded by the axionic entropy change. The framework suggests that
this intertwining is a general feature of topological magnets with
a proximate magnetic order parameter, and that it can be quantified
through the magnetoelectric response.

\section*{Author contributions}
A.M.U.\ conceived the study, performed all the density-functional
theory calculations, constructed and diagonalized the correlated
Hubbard--$J_1$--$J_2$ model, performed the tight-binding and Berry
curvature analysis, derived the axion electrodynamics and the
thermodynamic characterization, prepared all figures, wrote the
manuscript and handled the manuscript submission.

\section*{Data availability}
Data are available from the corresponding author on reasonable
request.

\section*{Acknowledgements}
Computational resources were provided by the Centre for High
Performance Computing (Lengau) under programme MATS0941. Funding
from the National Research Foundation of South Africa (Grant No.~115056) is gratefully
acknowledged.

\section*{Conflict of interest}
The authors declare no competing interests.

\end{document}